\documentclass[12pt]{article}
\usepackage{amsmath,epsfig}

\DeclareMathAlphabet\mathbb  {U}{msb}{m}{n}
\DeclareFontFamily{U}{msb}{} \DeclareFontShape{U}{msb}{m}{n}{
  <5> <6> <7> <8> <9> gen * msbm
  <10> <10.95> <12> <14.4> <17.28> <20.74> <24.88> msbm10
  }{}

\makeatletter
\def\section{\@startsection{section}{1}{\z@}{-3.25ex plus -1ex minus
             -.2ex}{1.5ex plus .2ex}{\normalfont\bfseries}}

\def\ps@hepth{\addtolength{\headheight}{12pt}
            \addtolength{\topmargin}{-15pt}
            \addtolength{\headsep}{15pt}
    \def\@oddhead{\hfil\small\begin{tabular}{r}
          \texttt{hep-th/0001182}\\ 
          UWThPh-2000-9\\
          SISSA/6/2000/FM
          \end{tabular}}
          \let\@evenhead\@oddhead
          \def\@oddfoot{\hfil\thepage\hfil}\let\@evenfoot\@oddfoot}
\makeatother

\renewenvironment{thebibliography}[1]
         {\section*{References}\frenchspacing\small
          \begin{list}{[\arabic{enumi}]}
         {\usecounter{enumi}\parsep=0pt\topsep 0pt
         \settowidth{\labelwidth}{[#1]}
         \leftmargin=\labelwidth\advance\leftmargin\labelsep
         \rightmargin=0pt\itemsep=0pt\sloppy}}{\end{list}}

\allowdisplaybreaks[3]

\begin{document}

\thispagestyle{hepth}

\begin{center}

$~$
\\[-2ex]
{\large \bfseries Renormalization of noncommutative 
Yang-Mills theories: \\[0.5ex] A simple example}
\vskip 2ex

\textsc{Harald Grosse}$^{a,}$\footnote{e-mail:
  \texttt{grosse@doppler.thp.univie.ac.at}}, 
\textsc{Thomas Krajewski}$^{b,}$\footnote{e-mail:
  \texttt{krajew@fm.sissa.it}} and 
\textsc{Raimar Wulkenhaar}$^{a,}$\footnote{e-mail:
  \texttt{raimar@doppler.thp.univie.ac.at}}
\vskip 1ex

{\small\itshape $^a\,$Institute for Theoretical Physics, University of
  Vienna\\ Boltzmanngasse 5, 1090 Wien, Austria
\\[1ex]
$^b\,$Scuola Internazionale Superiore di Studi Avanzati\\
via Beirut 4, 34014 Trieste, Italy}

\end{center}
\vskip 2ex

\begin{abstract}
  We prove by explicit calculation that Feynman graphs in
  noncommutative Yang-Mills theory made of repeated insertions into
  itself of arbitrarily many one-loop ghost propagator corrections are
  renormalizable by local counter\-terms. This provides a strong
  support for the renormalizability conjecture of that model.
\end{abstract}

\section{Introduction}

It is now commonly admitted that our current concepts about space and
time have to be changed when exploring space-time at a very small
scale. Indeed, one can show that it is impossible to locate a particle
with an arbitrarily small uncertainty when taking both into account the
principles of quantum mechanics and general relativity
\cite{doplicher}. Roughly speaking, one can say that measurements of
coordinates on space-time are subject to uncertainty relations, thus
ruining all geometrical concepts that have proved to be a guidance
principle in elaborating many physical theories.
 
\par

Following the previously alluded analogy with quantum mechanics, one
can try to solve this puzzle by assuming that the coordinates
themselves are noncommuting objects. Thus, the natural extension of
geometrical ideas to this new type of coordinates has been called
``noncommutative geometry''. Following even closer the ideas and
methods of quantum mechanics, we are led to assume that these
noncommutative coordinates are represented as a subalgebra of the
algebra of operators acting on a Hilbert space. This is the framework
of the theory pioneered by A. Connes \cite{connes}, which allows us
to make use of the powerful tools of functional analysis. Within this
framework, an analogue of gauge theory has been developed, even with
non trivial topological  properties, and it has already proved to be
useful in various areas of physics, ranging from the classical
description of the Higgs sector of the standard model (see
\cite{schucker} for a review) to recent ideas in string theory (see
\cite{witten} and references therein).

\par

This last example involves what we will call {\it NonCommutative 
Yang-Mills} (NCYM) theories in the sequel and can be thought of as a
generalization of non-abelian gauge theories, whose gauge symmetry and
interactions involves the noncommutative nature of the coordinates. A
first example of such a theory appeared almost ten years ago, when
Connes and Rieffel developed classical two dimensional Yang-Mills
theory on the noncommutative torus \cite{connesrieffel}. This idea has also
been generalized to higher dimensions \cite{spera}.

\par

This naturally raises the question of the quantization of such
theories, which has been tackled, at the one loop order, on the tori
in \cite{torus} and on noncommutative $\mathbb{R}^{D}$ in
\cite{martin} and \cite{sheikh}. Although these theories turned out to
be non local, i.e.\ their interacting vertices involve trigonometric
functions of the incoming momenta, it turns that the one-loop
behavior is quite similar to the standard non-abelian case. This
relies on an older work of Filk \cite{filk}, who proved that the
trigonometric factor of any planar graph (in a sense to be defined
below) does not involve trigonometric functions of the internal
momenta flowing into the loops, thus exhibiting the same divergence as
the standard theory.

\par

However, this is not true for non planar diagrams whose trigonometric
factor does involve a phase depending on the internal momenta.
Obviously such a phase softens the ultraviolet behavior of the
corresponding diagram and it has been conjectured by several authors
that such a diagram is in fact finite \cite{torus,bigatti,chepelev}.

\par

Nevertheless, it has been pointed out that this is not always the case
\cite{seiberg}. Indeed, if the non planar diagram contains some
special kind of non planar subdiagrams whose standard degree of
divergence is strictly greater that zero, which is the case in a
scalar field theory, the small momentum behavior of these diagrams
yields a new kind of infrared divergence intimately tied up with the non
locality.

\par

This short paper is devoted to a survey of this problem in the simple
case of multiloop corrections to the ghost propagator involving only
nested and disjoint subdivergent one loop corrections to the ghost
propagator. In the following section we shall briefly review the
problem raised in \cite{seiberg} and then we shall present an explicit
computation of the corresponding diagrams, postponing a complete probe
into the renormalization of NCYM theory to a future publication.

\section{Small momentum singularities induced by non planar diagrams}

Before entering into the details of NCYM theory, let us recall that
the noncommutative $\mathbb{R}^{D}$ is the algebra generated by $D$
hermitean elements $x_{\mu}$ with commutator $[x_{\mu},x_{\nu}]
=-2\mathrm{i} \theta_{\mu\nu}$, where $\theta_{\mu\nu}$ denotes a real
antisymmetric matrix which we will assume to be of maximal rank for
convenience.  Furthermore, one introduces Fourier modes
$U(k)=\mathrm{e}^{\mathrm{i} k\cdot x}$, with $k\cdot
x=k^{\mu}x_{\mu}$. We will always think of a smooth and at infinity
rapidly decreasing function as a Fourier transform
$$
f=\int \! d^{D}k\,f(k)U(k)~,
$$
where $k\mapsto f(k)$ is itself a smooth and rapidly decreasing
function on standard $\mathbb{R}^{D}$. The commutation relations of
the coordinates endow the algebra with the star product
\begin{equation}
f \star_{\theta}g :=\int\! d^{D}k \,d^{D}l\,f(k)g(l)\,U(k)U(l)~,
\end{equation}
which yields
$$
(f \star_{\theta} g)(k)=\int\! d^{D}l\, f(k{-}l) g(l)\,
\mathrm{e}^{\mathrm{i}\theta(k,l)}
$$
with $\theta(k,l)=\theta_{\mu\nu}k^{\mu}l^{\nu}$. Finally, this
algebra is equipped with the analogue of an integral defined as
\begin{equation}
\int\!f :=f(0)  
\end{equation}
and partial derivatives
\begin{equation}
\partial_{\mu}f :=\int\! d^{D}k\,\mathrm{i}k_{\mu}f(k)U(k)~,  
\end{equation}
which satisfy most of the properties of their commutative
counterparts: positivity and definiteness of the integral, Leibniz
rule, commutativity of partial derivatives and integration by part,
together with the tracial property of the integral
$$
\int\! f \star_{\theta} g = \int\! g \star_{\theta}f~, 
$$
which proves to be fundamental in the construction of gauge
invariant theories.

\par

At that point, two additional remarks are in order. First of all, let
us notice that the matrix $\theta_{\mu\nu}$ explicitly breaks Lorentz
invariance (or its euclidian counterpart), which is reduced to the 
transformations commuting with $\theta$, whereas translational
invariance is preserved. Furthermore, as $\theta_{\mu\nu}$ is
dimensionful, it also breaks scale invariance already at the classical 
level, for instance, in the case of four-dimensional NCYM theory or 
for a two-dimensional scalar field theory.

\par

From now on, one easily constructs scalar field theories, like
$\phi^{4}$, whose euclidian action is
\begin{equation}
S[\phi] = \int \Big( \frac{1}{2} \partial_\mu \phi \star_\theta 
\partial_\mu \phi +\frac{m^2}{2} \phi \star_\theta \phi +
\frac{g}{4!} \phi \star_\theta \phi \star_\theta \phi \star_\theta 
\phi\Big) ~,
\end{equation}
or the NCYM action
\begin{align}
S[A_{\mu}] &=-\frac{1}{4}\int\! F_{\mu\nu} \star_{\theta}F^{\mu\nu}  ~, 
\qquad \mbox{with}
\\
F_{\mu\nu} &=\partial_{\mu} A_{\nu}-\partial_{\nu}A_{\mu} 
+g( A_{\mu} \star_{\theta}A_{\nu}-A_{\nu} \star_{\theta}A_{\mu})~.
\notag
\end{align}
Because of the tracial properties of the integration, the latter
enjoys invariance under noncommutative gauge transformations
$$
\delta_{\lambda}A_{\mu}=g(\lambda  \star_{\theta} A_{\mu} 
-A_{\mu} \star_{\theta}\lambda) -\partial_{\mu}\lambda~.
$$

\par

Perturbative quantization of these theories is easily performed within
a formal functional integral point of view: The quadratic parts of the
actions are equal to their commutative counterparts, whereas the
interactions are non local and exhibit trigonometric functions of the
incoming momenta of the interaction vertices. Thus, the total
contribution of any Feynman diagram can be written as the product of a
rational function by a trigonometric function. Because trigonometric
functions are bounded, the standard rules of powercounting are
unchanged so that Weinberg's convergence theorem remains valid.

\par

It has been shown \cite{filk} that for planar diagrams the
trigonometric function is independent of the internal momenta so that
the Feynman integral reduces to the one encountered in a commutative
field theory. For non planar diagrams the situation is more involved
and we mainly have to distinguish two cases: whether the non planarity
results from crossing of internal lines (i.e.\ from the non planar
character of the amputed diagram), which we call type I diagrams, or
whether it solely comes from crossing of internal lines whith external
lines (type E). The latter case is much more tricky because the phase
vanishes when the corresponding external momenta satisfy some
particular relation.

\par 

The corresponding Feynman integral has been evaluated in \cite{seiberg}
within Schwinger's regularization scheme. It turns out that a type I
non planar diagram whose powercounting subdivergent non planar
diagrams are all of type I will be convergent, the corresponding
singularity when the Schwinger parameters goes to zero being removed. If
the diagram is of type E, but does not contain any subdivergence of
type E, it is non singular, except for some exceptional values of its
external momenta.

\par

Most of the trouble comes from the insertion of type E non planar
subdivergences. Indeed, the latter correspond to Feynman integrals of
the type
\begin{equation}
\int \!d^{nD}k\, R(k_{1},\dots,k_{n},p_{1},\dots, p_{N})\,
\mathrm{e}^{\mathrm{i}(\theta(k_{1},P_{1}) +\dots+\theta(k_{n},P_{n}))},
\end{equation}
where $k_{1},\dots,k_{n}$ are the independent internal momenta,
$p_{1},\dots,p_{N}$ are the external momenta and $P_{1},\dots,P_{n}$
are linear combinations of the external momenta.  The rational
function $R(k_{1},\dots,k_{n},p_{1},\dots, p_{N})$ is responsible for
the subdivergence.

\par

When all the $P_{i}$'s which couple to internal momenta belonging to
divergent integrals do not vanish, the corresponding integral can be
considered as finite, being regularized by the oscillatory factor.
Indeed, a computation with a cut-off introduced within Schwinger's
parametric formula yields such a finiteness. However, whenever one of the
$P_{i}$'s coupled to a divergent loop integral vanishes, then the
corresponding Feynman integral is just a usual divergent loop integral
and yields a singularity. When inserted into a larger diagram, this
could create some trouble when integrating over momenta approaching 
the subspace $P_{i}=0$. The question is how fast the divergence
appears compared with the smoothening property of the integration
measure. A quadratic divergence seems to destroy the renormalizability
\cite{seiberg} whereas a logarithmic divergence could be harmless. 
This conjecture is supported by our simple example.

\par

In a mathematically more satisfying manner, one can also consider such
an integral as a well defined distribution which is nothing but the
Fourier transform of the rational fraction $R$. However, we still have
to face a problem when inserting the distribution into a larger
Feynman diagram. In particular, the example described in
\cite{seiberg} corresponds to a distribution which is nothing but the
Feynman propagator and the infrared troubles when $p\rightarrow 0$ are
quite similar to the usual small $x$ singularities encountered in
QFT.

\par
  
Finally, let us point out that we did not encounter such a problem
when quantizing NCYM theory on a torus \cite{torus}. The construction
of the latter is similar to that of NCYM on $\mathbb{R}^{4}$ except
for the quantization of the momenta appearing in the Fourier
transform. As a consequence, there is no singularity when
$p\rightarrow 0$ since such a limit cannot be taken. However, we
noticed that a one loop non planar diagram with external momenta $p$
exhibits an extra UV singularity of the type $\delta(p)/\epsilon$.
Quite miraculously, all these singularities turn out to cancel,
leaving us with finite one loop renormalized correlators.

\par
 
In the next section, we shall evaluate the non planar contribution to
some of the one loop corrections to the ghost self energy using Bessel
functions, showing that they do not lead to any singularities when
inserted into larger diagrams. We shall use the Feynman rules for NCYM
theory without deriving them, the latter being obtained by replacing
the structure constant of non-abelian gauge theory $f_{abc}$ by
$2\mathrm{i}\sin \theta(p,q)$ \cite{martin}.

\section{A simple example. One-loop calculation}

Our goal is to compute the following 1-loop correction to the ghost
propagator in NCYM theory:
\begin{equation}
\parbox{40mm}{\begin{picture}(40,18)
\put(0,-8){\epsfig{file=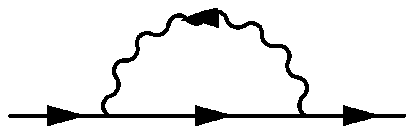}}
\put(5,7){\small$p$}
\put(35,7){\small$p$}
\put(17,7){\small$k{+}p$}
\put(19,16){\small$k$} 
              \end{picture}}
\end{equation}
Wavy lines represent gluons and straight lines ghosts. The Feynman
rules derived in \cite{martin} and (for the noncommutative torus) 
\cite{torus} lead to the integral 
\begin{align}
I_1 &= \int \!\! d^4k \; 4 g^2 \hbar ({-}p{-}k)^\mu 
\frac{(-1)}{k^2} \Big( \delta_{\mu\nu}-(1{-}\alpha)\frac{k_\mu
k_\nu}{k^2} \Big) \frac{(-1)}{(k{+}p)^2} (-p)^\nu 
\sin \theta(k,p) \,\sin \theta(-k,k{+}p) \notag
\\
&=  -4 g^2 \hbar \int \!\! d^4k \sin^2\theta(k,p) \Big( \frac{p^2 +\alpha
pk}{k^2(k+p)^2} - (1{-}\alpha) \frac{(pk)^2}{k^2 k^2 (k+p)^2}
\Big)~. 
\label{prop}
\end{align}
We work in a $D=4$ dimensional euclidian momentum space with metric 
$g_{\mu\nu}=\delta_{\mu\nu}$ and use obvious abbreviations such as
$pk=g_{\mu\nu}p^\mu k^\nu$. Using Feynman parameters
\begin{equation}
\frac{1}{A^r  B^s } = \frac{\Gamma(r{+}s)}{\Gamma(r) \Gamma(s)} 
\int_0^1 \frac{x^{r-1} (1{-}x)^{s-1} \,dx}{(Ax+B(1{-}x))^{r+s}}
\label{feynman}
\end{equation}
we obtain
\[
I_1 = -4 g^2 \hbar \int \!\!d^4k\, \sin^2 \theta(k,p) \int_0^1 \!\!
dx \Big( \frac{(p^2 {+}\alpha pk)}{(k^2 {+} 2pkx {+} p^2 x)^2} -
\frac{2(1{-}\alpha)(1{-}x)\,(pk)^2 }{(k^2 {+} 2pkx {+} p^2 x)^3} \Big)\,.
\]
In the denominator we write $qk$ instead of $pk$ so that we
reproduce the $k$'s in the numerator by differentiation with
respect to $q$:
\begin{align*}
I_1 = -4 g^2 \hbar \int \!\!d^4k\, \sin^2 \theta(k,p) \Big( p^2
\int_0^1 & \frac{dx}{(k^2 + 2qkx + p^2 x)^2}
- p^\mu \frac{\partial}{\partial q^\mu} \int_0^1 \frac{\alpha \,
dx}{2x (k^2 + 2qkx + p^2 x)}
\\
& \qquad - p^\mu p^\nu \frac{\partial}{\partial q^\mu}
\frac{\partial}{\partial q^\nu} \int_0^1
\frac{(1{-}\alpha)(1{-}x)\, dx}{4 x^2 (k^2 + 2qkx + p^2 x)}
\Big)\Big|_{q=p}~.
\end{align*}
Using $\frac{1}{A^n} = \frac{1}{\Gamma(n)} \int_0^\infty dt\,
t^{n-1}\, \mathrm{e}^{-tA}$ we rewrite the integral into
\begin{align*}
I_1 & = 4 g^2 \hbar \int_0^1 \!\!\! dx \int_0^\infty \!\!\! dt 
\Big( {-}p^2 t
+ \frac{\alpha}{2x} p^\mu \frac{\partial}{\partial q^\mu} +
\frac{(1{-}\alpha)(1{-}x)}{4 x^2} p^\mu p^\nu
\frac{\partial}{\partial q^\mu} \frac{\partial}{\partial q^\nu}
\Big) K[t,p,q,x] \Big|_{q=p}\,,
\\
& K[t,p,q,x] := \int d^4k\, \mathrm{e}^{-t(k^2 + 2qkx + p^2 x)}
\sin^2 \theta(k,p)~.
\end{align*}
Developing the sine into a Fourier series we obtain for the kernel 
\begin{align*}
K[t,p,q,x] =\int d^4k\, \big( \tfrac{1}{2} \mathrm{e}^{-t(k+qx)^2
- t(p^2 x - q^2 x^2)} 
& -\tfrac{1}{4} \mathrm{e}^{-t(k+qx+\mathrm{i}\theta(p)/t)^2 
- t(p^2 x - q^2 x^2) - 2 \mathrm{i}x \theta(p,q) -  p\circ p/t }
\\[-1ex]
&- \tfrac{1}{4} \mathrm{e}^{-t(k+qx-\mathrm{i}\theta(p)/t)^2 
- t(p^2 x - q^2 x^2) + 2 \mathrm{i}x \theta(p,q) - p\circ p/t }\big)~,
\end{align*}
where $\theta(p)^\mu := \theta^{\mu\alpha} p_\alpha$ and \cite{seiberg} 
$p\circ p := g_{\mu\nu} \theta^{\mu\alpha} p_\alpha \, \theta^{\nu\beta}
p_\beta \equiv (\theta(p))^2$. Then it is easy to perform the
Gaussian integration:
\begin{equation}
K[t,p,q,x] = \frac{\pi^2}{2 t^2} \big(\mathrm{e}^{- t(p^2 x - q^2
x^2)}  - \mathrm{e}^{- t(p^2 x - q^2 x^2)- p\circ p/t } \cos
2 x\theta(p,q) \big)~.
\end{equation}
We can now perform the differentiations with the following result:
\begin{align}
I_1 & = \pi^2 g^2 \hbar \int_0^1 \!\! dx \int_0^\infty \!\! dt \big(
\frac{1}{t} p^2 (3\alpha x {-} \alpha {-}1 {-}x) + 2 (p^2)^2
(1{-}\alpha) x^2(1{-}x) \big) \times \notag
\\[-1ex] 
& \hskip 16.5em \times
\big(\mathrm{e}^{- t p^2 x(1{-}x)} - \mathrm{e}^{- t p^2 x(1{-}x)-
p\circ p/t }\big) 
\\
&=p^2 \pi^2 g^2  \hbar \int_0^1 \!\!\! dx \int_0^\infty \!\!\! dt 
\big(\frac{1}{t} (3\alpha x {-}\alpha {-}1 {-}x) +
2(1{-}\alpha) x \big) \big(\mathrm{e}^{- t} -
\mathrm{e}^{- t - x(1{-}x) p^2 p\circ p/t }\big)\,. \notag
\end{align}
The integration over $t$ diverges logarithmically. We define the
projection $t^2_p$ onto the divergent part by
\begin{align}
t^2_p (I_1) &:= p^2 \pi^2 g^2 \hbar\int_0^1 \!\! dx \int_0^\infty
\frac{dt}{t} (3\alpha x {-}\alpha {-}1 {-}x)\mathrm{e}^{-t} 
= - p^2 \pi^2 g^2 \hbar \frac{3{-}\alpha}{2} \int_0^\infty
\frac{\mathrm{e}^{-t}\, dt}{t}\,. 
\label{div}
\end{align}
The projection to the convergent part is
\begin{align}
R_1 & := (1-t^2_p)(I_1) \label{conv}
\\[-1ex]
& =p^2 \pi^2 g^2 \hbar \Big( \! (1{-}\alpha) -\! \int_0^1 \!\!\! dx
\int_0^\infty \!\!\!\!\! dt \Big( \frac{3\alpha x {-} \alpha {-}1
{-}x}{t} + 2(1{-}\alpha) x \Big) \mathrm{e}^{- t - 
x(1{-}x) p^2 p\circ p/t }\Big). \notag
\end{align}
Instead of introducing a cutoff as in \cite{seiberg} we prefer such a
momentum subtraction before performing the divergent integral, in
analogy to the BPHZ scheme, as the notation $t_p^2$ for the projection
indicates. We believe this is advantageous for the renormalizability
proof to all orders based on Zimmermann's forest formula. Please
notice the difference between the power counting degree 2 entering the
forest formula and the actually only logarithmic divergence in
(\ref{div}). This is crucial for the insertion as subdivergences and
gives the reason why we will obtain local counterterms to all orders
whereas there are true quadratic divergences in the scalar theory in
\cite{seiberg}. It should be not difficult to extend such a
subtraction scheme to the entire NCYM theory. It is however important
to define $t_p^d$ as the projector onto the strictly divergent part of
an integral in order to produce local counterterms. This assumes one
can prove that all divergent integrations give rise to such local
terms, as we will do in this paper for repeated insertions of the
ghost propagator.

The $x$-integration in (\ref{conv}) yields $\int_0^1 dx\; x
\,\mathrm{e}^{-t-ax(1-x)/t
} = \frac{1}{2}
\int_0^1 dx\; \mathrm{e}^{-t-ax(1-x)/t}$ for any $a$ so that
\begin{align*}
R_1 & =p^2 \pi^2 g^2 \hbar \Big( (1{-}\alpha) + \int_0^1 \!\! dx
\int_0^\infty \!\!\! dt \big( \frac{3{-}\alpha}{2t} -
(1{-}\alpha) \big) \mathrm{e}^{- t - x(1{-}x) p^2
p\circ p/t }\Big).
\end{align*}
The $t$-integration leads to Bessel functions:
\begin{align}
R_1 &= p^2 \pi^2 g^2 \hbar \,(1{-}\alpha) \Big(1 - \int_0^1
\!\! dx \, 2 \sqrt{x(1{-}x)p^2p\circ p} \,K_1[2 
\sqrt{x(1{-}x)p^2p\circ p}]\Big)  \notag
\\[-1ex]
&+ p^2 \pi^2 g^2 \hbar \,(3{-}\alpha) \int_0^1 \!\! dx \,K_0[2 
\sqrt{x(1{-}x)p^2p\circ p}]~. 
\end{align}
The Bessel function $K_0[y]$ diverges logarithmically to $+\infty$ for
$y \to 0$ and converges exponentially to $0$ for $y \to +
\infty$. Hence, for any exponent $r>0$ there exists a number $c_r^0 >
0$ such that
\begin{equation}
K_0[y] \leq c^0_r/ y^r \qquad \forall ~0 < y < \infty~.
\label{K0}
\end{equation}
This will be proven algebraically in the Appendix. Graphically the
situation is sketched in Figure 1.%
\begin{table}[h] 
\epsfig{file=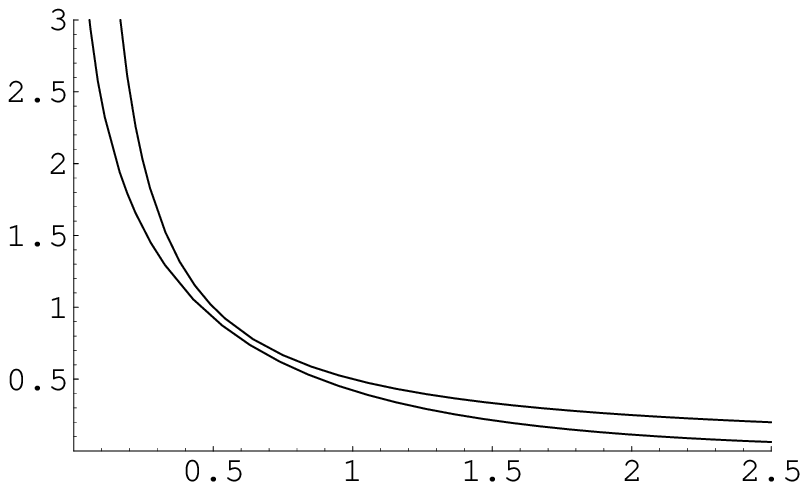,scale=0.8} \hfill
\raisebox{3.2cm}{\begin{minipage}[t]{75mm}
Figure~1: The Bessel function $K_0[y]$ (lower graph) is for any $0<y<\infty$
bounded by the function $c^0_r/y^r$ (upper graph). The situation is
shown for $r=1$ and $c^0_1 =\tfrac{1}{2}$.
\end{minipage}}
\end{table}
The function $y K_1[y]$ approaches~1 for $y \to 0$ and converges
exponentially to 0 for $y \to +\infty$. It is nevertheless convenient
to regard it as $K_0[y]$ before: For any exponent $r>0$ there exists a
number $c_r^1 > 0$ such that
\begin{equation}
y K_1[y] \leq c^1_r/ y^r \qquad \forall ~ 0 < y < \infty~.
\label{K1}
\end{equation}
The corresponding graphic is shown in Figure 2.%
\begin{table}[h] 
\epsfig{file=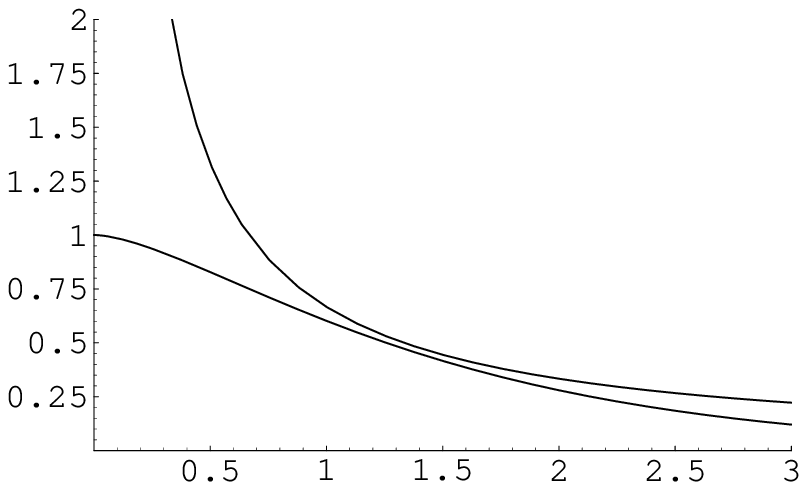,scale=0.8}\hfill 
\raisebox{3.2cm}{\begin{minipage}[t]{75mm}
Figure~2: 
The function $y K_1[y]$ (lower graph) is for any $0<y<\infty$
bounded by the function $c^1_r/y^r$ (upper graph). The situation is
shown for $r=1$ and $c^1_1 =\tfrac{2}{3}$.
\end{minipage}}
\end{table}
Now the $x$-integration is easy to perform. It converges for $0<r <1$,
which means that infrared divergences are absent (for non exceptional
momenta).

We restrict ourselves to the case where the rank of the tensor
$g_{\mu\nu} \theta^{\mu\alpha} \theta^{\nu\beta}$ equals the space-time
dimension (maximal noncommutativity). Then there exists some
parameter $m_P$ of dimension of a mass (the `Planck mass') such that 
\begin{equation}
p \circ p \geq \frac{p^2}{m_P^4} \qquad \Rightarrow\quad \sqrt{p^2 p
\circ p} \geq  \frac{p^2}{m_P^2}~.
\end{equation}
This yields the estimation 
\begin{equation}
R_1 =  p^2 \pi^2 g^2 \hbar \Big( (1{-}\alpha) + O\Big( P^1_r(\alpha) 
\Big(\frac{m_P^2}{p^2}\Big)^{\!r}\,\Big)\Big)~,
\end{equation}
where $P^n_r(\alpha)$ is a polynomial of homogeneous degree $n$ in
$(1{-}\alpha)$ and $(3{-}\alpha)$ with coefficients of order 1.

\section{Higher loop order calculation}

Now we insert $n$ of these 1-loop propagator corrections into a
propagator correction, giving an $n{+}1$ loop diagram: 
\begin{equation}
\parbox{100mm}{\begin{picture}(100,17)
\put(0,-3){\epsfig{file=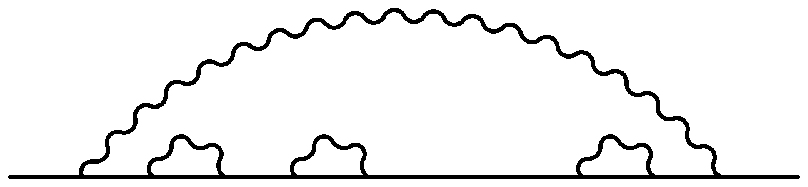}}
\put(30,5){\small$1$}
\put(45,5){\small$2$}
\put(68,5){\small$n$}
\put(60,17){\small$n{+}1$}
\put(55,2){$\dots$}
               \end{picture}}
\end{equation}
Using (\ref{prop}) and the Feynman rule \cite{torus} for the ghost
propagator, the corresponding integral is of order
\begin{align}
I_{n+1} =  - 4 g^2 \hbar(- \pi^2 g^2\hbar)^n 
\int \!\! d^4k \; & \sin^2\theta(k,p)
\big( \frac{p^2 +\alpha pk}{k^2(k+p)^2} - (1{-}\alpha)
\frac{(pk)^2}{k^2 k^2 (k+p)^2} \big)\times \notag
\\
& \hskip 2em \times \sum_{j=0}^n P^j_r(\alpha)
(1{-}\alpha)^{n-j} \Big(\frac{m_P^2}{(k+p)^2}\Big)^{jr} , 
\label{propn}
\end{align}
with $P_r^0(\alpha)=1$. Introduction of Feynman and Schwinger
parameters as before and use of $x \Gamma(x)=\Gamma(x{+}1)$ leads to
\begin{align}
I_{n+1} &=  -4 g^2 \hbar (- \pi^2 g^2\hbar)^n \sum_{j=0}^n
(1{+}jr)\,P^{j}_r (\alpha) \,(1{-}\alpha)^{n-j} m_P^{2jr} 
\int \!\! d^4k\, \sin^2\theta(k,p) 
\times \notag
\\*
& \hskip 5em \times \int_0^1 \!\!\! dx
\Big(\frac{x^{jr} (p^2 {+}\alpha pk)}{(k^2 {+} 2pkx {+} p^2 x)^{2+jr}} 
- \frac{(2{+}jr)(1{-}\alpha)(1{-}x)x^{jr} (pk)^2}{
(k^2 {+} 2pkx {+} p^2 x)^{3+jr}} \Big) \notag
\\
& = 4 g^2 \hbar (-\pi^2 g^2\hbar)^n \sum_{j=0}^n 
\frac{1}{\Gamma(1{+}jr)}\, P^{j}_r(\alpha) \,(1{-}\alpha)^{n-j} 
m_P^{2jr} \times \notag
\\*
& \times \int_0^1 \!\! dx \int_0^\infty \!\! dt \Big(
{-} x^{jr} t^{1+jr} p^2
+ \tfrac{1}{2} \alpha x^{jr-1} t^{jr} p^\mu
\frac{\partial}{\partial q^\mu} \notag
\\*[-1ex]
& \hskip 10em + \tfrac{1}{4} (1{-}\alpha)(1{-}x)x^{jr-2} t^{jr}
p^\mu p^\nu \frac{\partial}{\partial q^\mu}
\frac{\partial}{\partial q^\nu} \Big) K[t,p,q,x] \Big|_{q=p} \notag
\\
& = - (-\pi^2 g^2\hbar)^{n+1} \sum_{j=0}^n 
\frac{1}{\Gamma(1{+}jr)}\, P^{j}_r(\alpha) \,(1{-}\alpha)^{n-j} 
m_P^{2jr} \times \notag
\\*[-1ex]
& \times \int_0^1 \!\! dx \int_0^\infty \!\! dt \Big(
(3\alpha x {-}x{-}\alpha{-}1) x^{jr} t^{jr-1} p^2 
+ 2(1{-}\alpha)(1{-}x)x^{jr+2} t^{jr} (p^2)^2 \Big) \times \notag 
\\*
& \hskip 16em \times
\big(\mathrm{e}^{- t p^2 x(1{-}x)} - \mathrm{e}^{- t p^2 x(1{-}x)-
p \circ p/t }\big) \notag 
\\
& = -(-\pi^2 g^2\hbar)^{n+1} p^2 \sum_{j=0}^n 
\frac{1}{\Gamma(1{+}jr)}\, P^{j}_r(\alpha) \,(1{-}\alpha)^{n-j} 
\Big(\frac{m_P^2}{p^2}\Big)^{jr} \int_0^1 \!\! \frac{dx}{(1{-}x)^{jr}} 
\times \notag
\\*[-1ex]
& \times \int_0^\infty \!\! dt \big(
(3\alpha x{-}x{-}\alpha{-}1) t^{jr-1} + 2(1{-}\alpha)x t^{jr}\big) 
\big(\mathrm{e}^{- t} - \mathrm{e}^{-t - x(1{-}x) p^2 p \circ p /t}
\big)\,. \raisetag{8ex}
\end{align}
The only divergent integral is for $j=0$ the projection 
\begin{align}
t^2_p(I_{n+1}) & :=  - (-\pi^2 g^2\hbar)^{n+1} p^2 (1{-}\alpha)^n
\int_0^1 \!\! dx \int_0^\infty \! \frac{dt}{t} 
(3\alpha x{-}x{-}\alpha{-}1) \mathrm{e}^{- t} \notag
\\
=&  (-\pi^2 g^2\hbar)^{n+1} p^2 (1{-}\alpha)^n \, \frac{3{-}\alpha}{2}
\int_0^\infty \! \frac{dt}{t} \mathrm{e}^{- t}~.
\label{tn}
\end{align}
The convergent part can be evaluated to
\begin{align}
R_{n+1} & = (1-t_p^2)(I_{n+1} ) \notag \\
& = - p^2 (-\pi^2 g^2\hbar)^{n+1} (1{-}\alpha)^{n+1} 
\Big(1 - \int_0^1\!\! dx \, 
2 \sqrt{x(1{-}x) p^2 p\circ p} \,
K_1[2 \sqrt{x(1{-}x) p^2 p\circ p}]\Big) \notag
\\
& - p^2 (-\pi^2 g^2 \hbar)^{n+1} (1{-}\alpha)^n (3{-}\alpha) 
\int_0^1 \!\! dx \,K_0[2 \sqrt{x(1{-}x)p^2 p\circ p}] \notag
\\
& + p^2 (-\pi^2 g^2\hbar)^{n+1} \sum_{j=1}^n 
\frac{P^{j}_r(\alpha)}{jr(2{-}jr)} \,(1{-}\alpha)^{n-j} (3{-}\alpha) 
\Big(\frac{m_P^2}{p^2}\Big)^{jr} 
\\
& + p^2 (-\pi^2 g^2\hbar)^{n+1} \sum_{j=1}^n 
\frac{1}{\Gamma(1{+}jr)}\, P^{j}_r(\alpha) \,(1{-}\alpha)^{n-j} 
\Big(\frac{m_P^2}{p^2}\Big)^{jr} \times \notag
\\[-1ex]
& \times \int_0^1 \!\! \frac{dx}{(1{-}x)^{jr}} \int_0^\infty \!\!\! dt 
\big( (3\alpha x{-}x{-}\alpha{-}1) t^{jr-1} + 2(1{-}\alpha)x t^{jr}\big) 
\mathrm{e}^{-t - x(1{-}x)p^2 p\circ p /t}\,. \notag
\end{align}
The last two lines range from zero (for $p=\infty$) to minus the value of
the third last line for $p=0$. Thus, in our estimation we have to
neglect the last two lines. The remaining integrals over $K_0$ and
$K_1$ are familiar to us, see Figures 1 and 2, and we choose the
essential exponents in (\ref{K0}) and (\ref{K1}) to be
$(n{+}1)r$ instead of $r$. Then we arrive at
\begin{equation}
R_{n+1} = - p^2 (-\pi^2 g^2\hbar)^{n+1} \Big( (1{-}\alpha)^{n+1}
+ O\Big( \sum_{j=1}^{n+1} P^{n+1}_{r,j}(\alpha)
\Big(\frac{m_P^2}{p^2}\Big)^{\!jr}\, \Big)\Big)~.
\label{Rn}
\end{equation}
But this was precisely our starting point we inserted into
(\ref{prop}) to obtain (\ref{propn}). Hence, (\ref{Rn}) 
provides the structure of any renormalized $n{+}1$ loop graph made of
ghost propagator corrections. The counterterm of such an $n$-loop graph
is given by (\ref{tn}), and we see explicitly that the Feynman graphs
made of nested 1-loop ghost propagator corrections
\begin{equation}
\parbox{100mm}{\begin{picture}(100,16)
\put(0,0){\epsfig{file=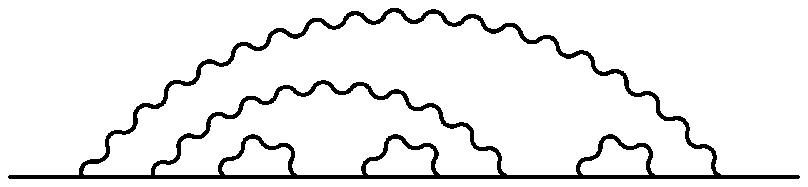,scale=0.8}}\end{picture}}
\end{equation}
are renormalized by local counterterms for any order in $\hbar$. Here,
locality means that the momentum dependence of the counterterm and the
kinetic part of the ghost action are identical. In
order to renormalize an $n$-loop graph (to avoid infrared divergences)
the critical exponent has to be chosen $0 < r < 1/n$.

The essential step in this proof was the observation that the
noncommutative Feynman graphs under consideration
evaluate to Bessel functions, which can be estimated by a power
law. It seems plausible that any Feynman graph of noncommutative
Yang-Mills theory evaluates to Bessel functions, and applying the same
techniques it should be possible to show that local counterterms
suffice to renormalize this model.

\section*{Appendix: Proof of Eq.\ (\ref{K0})}

We prove that for each $r>0$ there is a number $c_r>0$ such that 
\[
C_0(x):=\frac{c_r}{x^r} \geq K_0(x)\qquad \forall~0<x<\infty~.
\]
The Bessel function $K_0(x)$ is one of the two solutions of the 
differential equation
\begin{equation}
x K_0'' + K_0' -xK_0 =0~,\qquad 0<x<\infty~.
\label{bessel}
\end{equation}
It is however more convenient to consider the function 
\[
K(x):=\sqrt{x} K_0(x)~, \qquad
K'' + \frac{1 -4x^2}{4x^2}K =0~,
\]
and compare it with
\[
C(x):=\sqrt{x} C_0(x)~,\qquad 
C'' + \frac{1-4 r^2}{4 x^2} C =0~.
\]
The derivative of the Wronskian $W(K,C):= K'C-C'K$ is
\[
W' = \frac{x^2 - r^2}{x^2} KC~,\qquad \begin{array}{ll} W'> 0 & 
\mbox{ for } x>r \\
W'< 0 & \mbox{ for } x<r  \end{array}
\]
The asymptotic development shows that for the solution $K_0$ of
(\ref{bessel}) one has $W(x)<0$ for $x\to \infty$ and
$W(x)>0$ for $x\to0$. Therefore, there is only one zero of the
Wronskian, at $x=x_r$, as illustrated in Figure 3.%
\begin{table}[h]
\epsfig{file=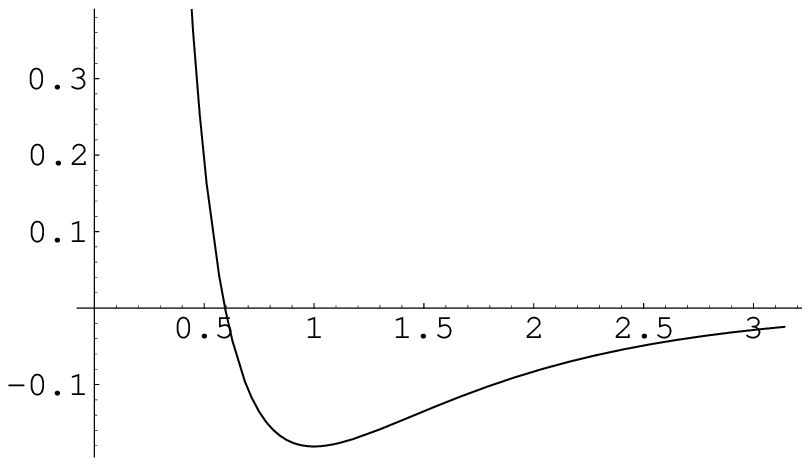,scale=0.8}\hfill
\raisebox{2.8cm}{\begin{minipage}[t]{77mm}
Figure~3: The Wronskian $W(K,C)$ has an extremum for $x=r$ and a zero
for $x=x_r$. The situation is shown for $r=1$ and $c_r =1$. 
\end{minipage}}
\end{table}

\noindent
We choose the normalization 
\[
K_0(x_r) = c_r x_r^{-r}~,\qquad K_0'(x_r) = -r c_r x_r^{-r-1}
\]
so that $W(x_r)=0$ and $K(x_r)=C(x_r)$. Then we can integrate
\begin{align*}
& x>x_r ~\Rightarrow~ W(x)<0 : \quad  
&\int_{x_r}^x \frac{K'(x)}{K(x)} dx < 
\int_{x_r}^x \frac{C'(x)}{C(x)} dx \quad & \Rightarrow~ K(x) < C(x)
\\
& x<x_r ~\Rightarrow~ W(x)>0 : \quad  
&\int^{x_r}_x \frac{K'(x)}{K(x)} dx > 
\int^{x_r}_x \frac{C'(x)}{C(x)} dx \quad & \Rightarrow~ K(x) < C(x)
\end{align*}
This finishes the proof.

\end{document}